\documentclass{aa}
\usepackage[varg]{txfonts}
\usepackage{graphicx,lscape, soul, color}
\bibpunct{(}{)}{;}{a}{}{,}

\begin{document}

\title{Effects of M dwarf magnetic fields on potentially habitable planets}

\author{A.~A.~Vidotto\inst{1}
\and M.~Jardine\inst{1}
\and J. Morin\inst{2}
\and J.-F. Donati\inst{3}
\and P.~Lang\inst{1}
\and A.~J.~B.~Russell\inst{4}}
\institute{SUPA, School of Physics and Astronomy, University of St Andrews, North Haugh, KY16 9SS, St Andrews, UK\and
Institut f\"ur Astrophysik, Georg-August-Universit\"at, Friedrich-Hund-Platz 1, D-37077, Goettingen, Germany \and
Laboratoire d'Astrophysique, Observatoire Midi-Pir\'en\'ees, 14 Av.~E.~Belin, F-31400, Toulouse, France\and
SUPA, School of Physics and Astronomy, University of Glasgow, University Avenue, G12 8QQ, Glasgow, UK}

\date{Received date /
Accepted date }

\abstract{{{We investigate the effect of the magnetic fields of M dwarf (dM) stars on potentially habitable Earth-like planets. These fields can reduce the size of planetary magnetospheres to such an extent that a significant fraction of the planet's atmosphere may be exposed to erosion by the stellar wind.}
{We used a sample of 15 active dM stars, for which surface magnetic-field maps were reconstructed, to determine the magnetic pressure at the planet orbit and hence the {\it largest} size of its magnetosphere, which would only be decreased by considering the stellar wind. Our method provides a fast means to assess which planets are most affected by the stellar magnetic field, which can be used as a first study to be followed by more sophisticated models.} 
{We show that hypothetical Earth-like planets with similar terrestrial magnetisation ($\sim 1~$G) orbiting at the inner (outer) edge of the habitable zone of these stars would present  magnetospheres that extend at most up to $6$ ($11.7$) planetary radii. To be able to sustain an Earth-sized magnetosphere, with the exception of only a few cases, the terrestrial planet would either (1) need to orbit significantly farther out than the traditional limits of the habitable zone;  or else, (2) if it were orbiting within the habitable zone, it would require at least a magnetic field ranging from a few G to up to a few thousand G. 
By assuming a magnetospheric size that is more appropriate for the young-Earth ($3.4$~Gyr ago), the required planetary magnetic fields are one order of magnitude weaker. 
However, in this case, the polar-cap area of the planet, which is unprotected from transport of particles to/from interplanetary space, is twice as large. At present, we do not know how small the smallest area of the planetary surface is that could be exposed and would still not affect the potential for formation and development of life in a planet. 
As the star becomes older and, therefore, its rotation rate and magnetic field reduce, the interplanetary magnetic pressure  decreases and the magnetosphere of planets probably expands. Using an empirically derived rotation-activity/magnetism relation, we provide an analytical expression for estimating the shortest stellar rotation period for which an Earth-analogue in the habitable zone could sustain an Earth-sized magnetosphere. We find that the required rotation rate of the early- and mid-dM stars (with periods $\gtrsim 37$ -- $202$~days) is slower than the solar one, and even slower for the late-dM stars ($\gtrsim 63$ -- $263$~days).  }
Planets orbiting in the habitable zone of dM stars that rotate faster than this have smaller magnetospheric sizes than that of the Earth magnetosphere. Because many late-dM stars are fast rotators, conditions for terrestrial planets to harbour Earth-sized magnetospheres are more easily achieved for planets orbiting slowly rotating early- and mid-dM stars.}} 

\maketitle

\section{INTRODUCTION}\label{sec.intro}
M dwarf (dM) stars have been the prime targets for terrestrial-planet searches for two reasons: (1) because they are inherently low-luminosity objects, they provide a good contrast to detect smaller-radius planets in transit searches and (2) because they are low-mass objects, the reflex motion induced by a terrestrial planet is within reach of current spectrograph sensitivities in radial velocity searches. Another interesting aspect of dM stars is that the region where life could potentially develop (the habitable zone, HZ) is located significantly closer to dM stars than it is for solar-type stars \citep{1993Icar..101..108K, 2007A&A...476.1373S}. Based on incident stellar flux arguments, a terrestrial planet orbiting inside this region should be able to retain liquid water at its surface. The combination of a HZ that is closer to the star and the technologies currently adopted in exoplanet searches make dM stars the prime targets for detecting terrestrial planets in the potentially life-bearing region around the star.  

However, in addition to the retention of liquid water, other factors may be important in assessing the potential for a planet to harbour life \citep[see comprehensive reviews by][and references therein]{2007AsBio...7...85S,2007AsBio...7...30T,2009A&ARv..17..181Lb}. For example, there has been a great deal of work spent in investigating how planets orbiting dM stars may be affected by stellar ejecta. \citet{2007AsBio...7..167K} and \citet{2007AsBio...7..185L} investigated how coronal mass ejections (CMEs) might affect potentially habitable planets, while \citet{2005AsBio...5..587G,2009Icar..199..526G} focused on the impact of stellar (and galactic) cosmic rays on such planets. These works suggested that because particle (as well as X-ray and extreme-ultraviolet radiation, \citealt{2007AsBio...7..185L}) exposure can strongly impact the atmospheres of terrestrial-type planets in these HZs, the HZ extent might be  narrower then the traditional definition. Quiescent stellar outflows can also be harmful for the creation and development of life, because a strong stellar wind alone can erode an unprotected planetary atmosphere on a short time scale \citep{2010Icar..210..539Z}. \citet{2011MNRAS.412..351V} showed that fast-rotating dM stars may host winds that are considerably different from the solar wind, and that when such winds interact with a planet orbiting in the HZ, erosion of the planet atmosphere is expected unless the planet is protected by a more intense magnetic field than that of the Earth. The presence of a relatively strong planetary magnetic field is, therefore, very likely to play a significant role in planetary habitability. 

Theoretically, the efficiency of a planetary dynamo is related to its interior structure: it should possess an electrically conducting region and a convective flow \citep[e.g.,][]{2003E&PSL.208....1S}. Contrary to past beliefs, recent studies indicate that the planetary field strength is independent of rotation rate, which instead plays a role in the geometry of the generated magnetic field (dipolar or multipolar). Numerical simulations have shown that the fraction of dipolar field depends on the local Rossby number\footnote{The empirical Rossby number ${\rm Ro}$ is defined as the rotation period of the star/planet over the characteristic convective turnover time scale. The local Rossby number is ${\rm Ro}_l = u_{\rm conv} /(\Omega l)$, where $u_{\rm conv}$ is the rms velocity of the convective motion, $\Omega$ the rotation rate and $l$ is the typical flow lengthscale \citep{2013A&A...549L...5G}.}, ${\rm Ro}_l$, where for ${\rm Ro}_l \lesssim 0.1$, the planet is believed to be in the dipolar-dominated regime \citep[e.g.,][]{2006GeoJI.166...97C}. Because $\rm{Ro}_l$ is almost linearly proportional to the planetary rotation period\footnote{Keplerian orbital periods of planets in the HZ of dM stars can range between $\sim 2$ and $114$~d, more commonly $\gtrsim 10$~d for a combination of many stellar masses and orbital distances.}, a tidally locked Earth-like planet in the HZ of dM stars would probably have ${\rm Ro}_l>1$ and, therefore, a weak dipolar field \citep{2012Icar..217...88Z}. In that case, these planets would lack a protective magnetic field, potentially losing a significant fraction of their atmospheres \citep{2013arXiv1304.2909Z}. 

The magnetic field is a quantity that has not yet been directly observed on extrasolar planets, in spite of many attempts to detect planetary radio emission \citep[e.g.,][]{2000ApJ...545.1058B,2004ApJ...612..511L,2013ApJ...762...34H,2013arXiv1302.4612L}. If confirmed, the technique proposed by \citet{2010ApJ...722L.168V}, based on near-UV transit observations, should provide a useful tool in determining planetary magnetic field intensities for hot-Jupiter transiting systems, but may be more limited in the case of terrestrial planets orbiting dM stars \citep{2011AN....332.1055V}. 

In steady state, the extent of a planet's magnetosphere is determined by force balance at the boundary between the stellar coronal plasma and the planetary plasma, a method that has been used by the solar system community for several decades \citep[e.g.,][]{1930Natur.126..129C}. For the planets in the solar system, this is often reduced to a pressure balance at the dayside, the most significant contribution to the external (stellar) wind pressure being the solar wind ram pressure \citep[e.g.][]{1995isp..book.....K}. However, for planets orbiting stars that are significantly more magnetised than the Sun or/and are located at close distances, the stellar magnetic pressure may play an important role in setting the magnetospheric limits \citep{2004ApJ...602L..53I,2006JGRA..111.6203Z,2008MNRAS.389.1233L,2009A&A...505..339L,2009ApJ...703.1734V,2010ApJ...720.1262V,2012MNRAS.423.3285V,2011MNRAS.412..351V,2011JGRA..116.1217S,2012ApJ...744...70K,2013ApJ...765L..25B}. The extent of the magnetosphere of planets orbiting in the HZ of dM stars have been investigated by other authors \citep[e.g.][]{2005AsBio...5..587G,2009Icar..199..526G, 2007AsBio...7..167K, 2007AsBio...7..185L,2011MNRAS.412..351V,2011AN....332.1055V}, but, to the best of our knowledge, a detailed analysis of the particular contribution of the stellar magnetic field remains to be made. Part of this limitation is justified by the fact that it was only recently that the large-scale magnetic field of dM stars was reconstructed for the first time \citep{2006Sci...311..633D}. Since then, new observations showed that dM stars can harbour magnetic fields that are quite different from the solar one both in intensity and topology \citep{2008MNRAS.390..545D,2008MNRAS.384...77M,2008MNRAS.390..567M,2010MNRAS.407.2269M,2009ApJ...704.1721P}. In particular, because of the close location of their HZ, planets orbiting in this region interact with significantly more intense interplanetary magnetic field than does the Earth.

In the present work, we quantitatively evaluate the sizes of planetary magnetospheres resulting from the pressure exerted by the intense stellar magnetic fields found around M-dwarf stars. Our approach only invokes a stellar magnetic field, neglecting effects such as dynamic pressures. We show in Section~\ref{sec.model} that our approach provides a useful {\it upper} limit for magnetospheric sizes -- magnetospheres will tend to be even smaller if the ram pressures of stellar winds and CMEs are accounted for. Our technique also has the advantage that it can be employed by anyone in the astronomical community, without requiring sophisticated magnetohydrodynamic numerical simulations.
It provides, therefore, a fast means to assess which planets are most affected by the stellar magnetic field. Section~\ref{sec.results} investigates the sizes of the magnetospheres of planets orbiting a sample of 15 active dM stars, for which surface magnetic field maps were reconstructed. Section~\ref{sec.results2} focuses on the weakly active dM stars. Our summary and final remarks are presented in Section~\ref{sec.conc}.

\section{Model}\label{sec.model}
\subsection{Interaction between the planet and the corona of its host star}\label{sec.pl_mag}
The magnetosphere of a planet carves a hole in the stellar coronal plasma, because it deflects the coronal material around it. At a characteristic distance $r_M$ from the planet centre (i.e., the characteristic magnetospheric size), pressure balance implies that the total pressure from the stellar plasma surrounding the planet $P_{\rm ext} (R_{\rm orb})$ balances the total pressure from the planetary plasma $P_{\rm pl} (r_M)$, where  $R_{\rm orb}$ is the orbital radius of the planet. 

On the planetary side, pressures resulting from thermal motions, mass loss (atmospheric escape), and magnetic effects  can all contribute to the pressure equilibrium. The Earth magnetosphere is located at  $r_M\simeq 10$ -- $15~R_\oplus$ \citep{1992AREPS..20..289B}, where  planetary thermal pressure can safely be neglected. The Earth also does not have a significant atmospheric escape \citep[$\lesssim$ a few kg/s,][]{1983RvGSP..21...75F}. Mass loss from planets can be significant, however, especially for close-in gas giant planets orbiting solar-type stars (or hotter), as a consequence of high stellar irradiation \citep{2007A&A...461.1185L,2009ApJ...693...23M,2011A&A...529A.136E, 2012MNRAS.425.2931O}. Here, we assume that ram pressure due to atmospheric escape of terrestrial planets (with limited gas reservoir) orbiting in the HZ of dM stars is negligible. Therefore, the main factor contributing in the pressure balance equation at the planetary side for an Earth-like planet with a similar-to-terrestrial magnetisation is its own magnetic pressure  
\begin{equation}\label{eq.pint}
P_{\rm pl}(r_M) \simeq \frac{[B_{\rm pl}(r_M)]^2}{8\pi} ,
\end{equation}
where $B_{\rm pl}(r_M)$ is the intensity of the planet magnetic field at a distance $r_M$ from the planet centre. 

On the stellar side, factors such as the thermal pressure, magnetic pressure ($P_{B,\star}$) and the ram pressure resulting from the relative motion between the planet and the coronal material can all contribute to setting the pressure equilibrium. 
The stellar-wind properties determine, for example, whether Earth-type magnetospheres surrounded by bow shocks or Ganymede-type magnetospheres with Alfv\'en wings are formed \citep{2004ApJ...602L..53I, 2007P&SS...55..598Z,2011ApJ...729..116K,2011JGRA..116.1217S,2013A&A...552A.119S}. To quantitatively evaluate this, the stellar wind density, temperature, magnetic field, and the relative velocity of the planet are required together with planetary magnetic characteristics. Using an isothermal stellar wind \citep{1958ApJ...128..664P} with typical coronal temperatures ($10^6$~K), \citet{2011AN....332.1055V} showed that the stellar wind of a dM star becomes super-sonic at about $5$ stellar radii ($\sim 0.008$~au) for a typical M4.0 star, while for a typical M1.5 star, the sonic point lies at about $8$ stellar radii ($\sim 0.018$~au). This implies that planets orbiting at the HZ of dM stars interact with super-sonic stellar winds. A more complete stellar wind model, which incorporates the complex stellar magnetic field topology and rotation, is nevertheless essential to evaluate whether the stellar-wind interaction is super-Alfv\'enic (giving rise to  bow shocks surrounding the magnetospheres of planets) or whether the interaction is sub-Alfv\'enic and Alfv\'en wings are formed \citep[e.g.,][]{2013A&A...552A.119S}.  Three-dimensional simulations of winds of dM stars suggest that the stellar wind is super-Alfv\'enic at the HZ \citep{2011MNRAS.412..351V}, a result that is also obtained in the analytical models of \citet{2013A&A...552A.119S}. This implies that terrestrial-type magnetospheres surrounded by bow shocks are expected to be present around magnetised planets orbiting dM stars at their HZs. Unfortunately, we have little information on the magnitude of the thermal and ram pressures of the stellar winds of dM stars, because the low-density, optically thin winds of these stars prevent the observation of traditional mass-loss signatures, such as P Cygni profiles. As a consequence, estimates of mass-loss rates from dM stars (and therefore, of their stellar wind densities and velocities) are still debatable (see discussion in \S1 of \citealt{2011MNRAS.412..351V}). In view of this uncertainty, we take
\begin{equation}\label{eq.pextmin}
P_{\rm ext} (R_{\rm orb})  \geq P_{\rm ext}^{\rm min} (R_{\rm orb}) = P_{B,\star} (R_{\rm orb}) = \frac{[B_\star(R_{\rm orb})]^2}{8\pi} \, ,
\end{equation}
where $P_{\rm ext}^{\rm min} (R_{\rm orb})$ is the {\it lowest external pressure that is exerted on the planet}. {Adding the stellar wind thermal and ram pressures will increase the upstream external pressure beyond this minimum (see \S\ref{subsec.part1} for estimates on how the consideration of stellar wind ram pressure affects our results). Therefore, because $P_{\rm ext} (R_{\rm orb})=P_{\rm pl}(r_M)$, from Eqs.~(\ref{eq.pint}) and (\ref{eq.pextmin}), we have }
\begin{equation}\label{eq.disequilibrium}
 \frac{[B_{\rm pl}(r_M)]^2}{8\pi} \geq \frac{[B_\star(R_{\rm orb})]^2}{8\pi} ,
\end{equation}
{i.e., the magnetic pressure of the planet at a distance $r_M$ from its centre is greater than the pressure of the stellar magnetic field at the orbital radius of the planet.}

Assuming a dipolar planetary magnetic field, we have that $B_{\rm pl} (r) = \frac12 B_{p,0} (r_p/r)^3$ in the magnetic equatorial plane, where $r_p$ is the planetary radius, $r$ is the radial coordinate centred at the planet and $B_{p,0}$ is the surface magnetic field at the pole. Therefore, the {\it upper} limit on the magnetospheric size can be calculated as 
\begin{equation}\label{eq.rm}
 \frac{r_M^{\max}}{r_p} = \left(\frac{B_{p,0}/2}{B_\star(R_{\rm orb})}\right)^{1/3} .
\end{equation}

In the solar system, the magnetised planets interact with a super-magnetosonic solar wind that deforms each magnetosphere, compressing it on the upwind side, elongating it on the downwind side, and forming an upstream bow shock. It is therefore common practice in the solar system community to define $r_M$ as the position of the magnetospause at the nose, taking ram pressure as the dominant external pressure at this location. Eq.~(\ref{eq.disequilibrium}) provides an estimate of a size that is more appropriate for characterising the distance to the flanks of the terrestrial magnetosphere (where the effect of the ram pressure of the solar wind is smaller). We therefore stress that $r_M$ here refers to a characteristic distance to the point where magnetic pressure equilibrium exists, which places an {\it upper} limit on the magnetospheric size, with the size at the nose being smaller if the stellar wind ram pressure is also taken into account. 

We note that the relative orientation of the stellar magnetic field with respect to the orientation of the planetary magnetic moment plays an important role in shaping the open-field-line region on the planet \citep[e.g.,][]{2004ApJ...602L..53I,2006JGRA..111.6203Z, 2011ApJ...729..116K,2011JGRA..116.1217S,2013A&A...552A.119S}. Our purely magnetic pressure balance (Eq.~(\ref{eq.disequilibrium})) reduces to a result similar to what one would have obtained by solving for the position of magnetic nulls in a superposition of a planetary dipole and uniform interplanetary magnetic field when such a field is parallel to the planet's magnetic moment \citep[e.g.,][]{1961PhRvL...6...47D}. Note, however, that certain configurations can result in closed magnetospheres, i.e., without a polar-cap area. Here, $r_M$ is taken as an upper limit on the magnetospheric size when the magnetosphere is in its widest open configuration, as will occur regularly since the complex  magnetic-field topology of the stars in our sample presents non-uniform directions and strengths (cf.~Section~\ref{sec.maps}). In addition, for the complex topology of the stellar magnetic field and of the yet poorly known characteristics of the planetary magnetic field, such an alignment is also a function of the obliquity of the planetary orbit. We note that these quantities will vary from system to system. 

In the present work, only intrinsic planetary magnetic fields are considered. The formation of induced magnetic fields could increase the magnetospheric size estimated in Eq.~(\ref{eq.rm}). It is believed that Jupiter-like planets with additional plasma sources, such as from active outgassing moons or mass loss from the planet itself, can harbour induced magnetic fields produced by magnetodisc ring currents \citep{2011MNRAS.414.2125N,2012MNRAS.427L..75N,2012ApJ...744...70K}. \citet{2012ApJ...744...70K} estimated that these fields can dominate over the contribution of an intrinsic magnetic dipole in the case of hot-Jupiters with strong mass-loss that orbit solar-like stars. They showed that this can result in magnetospheres that are more extended by a factor of 1.4 -- 1.7 than those traditionally estimated by considering only the intrinsic planetary dipole. Magnetic fields may also be induced if there is a direct interaction of the planet's ionosphere with the stellar wind, in a similar way as the induced Venusian magnetosphere \citep{1970P&SS...18.1281S,2008P&SS...56..790Z}.  To the best of our knowledge, the currently available studies of induced magnetic fields in exoplanets have been concentrated on gaseous planets, which should naturally present plasma sources within their magnetospheres (e.g., outgassing moons or enhanced mass loss). \citet{2009ApJ...703..905T} suggested that super-Earth-type planets orbiting in the HZ of dM stars present atmospheres stable against thermal escape. If thermal escape is the only source of plasma within the planetary magnetosphere, \citeauthor{2009ApJ...703..905T}'s  result seems to indicate that it is probably difficult to generate induced fields in these rocky planets due to their limited gas reservoir. However, these planets might contain other sources of plasma, such as from outgassing moons or escape due to non-thermal processes. It is, therefore, still unclear if intense induced magnetic fields would be generated and maintained in terrestrial rocky planets.

\subsection{Prescription for the stellar magnetic field}\label{sec.maps}
The goal of this paper is to investigate the contribution of the stellar magnetic field on setting the sizes of magnetospheres of planets orbiting dM stars. Therefore, to relate the observed surface stellar magnetic field to the magnetic fields expected at the orbits of planets, we employed magnetic-field extrapolations of the observationally reconstructed large-scale, surface magnetic maps. 

\subsubsection{Star sample} 
The surface magnetic maps adopted here were reconstructed using Zeeman-Doppler imaging (ZDI), a tomographic imaging technique \citep[e.g.,][]{1997A&A...326.1135D}. Using ZDI, one can reconstruct the large-scale magnetic field (intensity and orientation) at the surface of the star from a series of circular polarisation spectra. In this work, we concentrated on dM stars and used the maps that were published in \citet{2008MNRAS.390..545D} and \citet{2008MNRAS.390..567M,2010MNRAS.407.2269M}. Our sample consists of $15$ stars with masses ranging from $0.1$ to $0.75~M_\odot$, spanning spectral types M0 -- M6. We note that the stars in our sample harbour much stronger large-scale fields than the large-scale (non-active region) surface fields in the Sun ($\sim 1$~G). Table~\ref{table_B} shows some properties of the entire sample of dM stars considered. Data from the first eight columns are taken from \citet{2008MNRAS.390..545D} and \citet{2008MNRAS.390..567M,2010MNRAS.407.2269M}. Effective temperatures and bolometric luminosities are derived from NextGen models \citep{1998A&A...337..403B}. The remaining columns are results of our model (Section~\ref{sec.results}).

\begin{sidewaystable}
\caption{Sample of dM stars. The columns are: the star name, the observation year, the stellar spectral type, mass $M_*$, radius $R_*$, rotation period $P_{\rm rot}$, Rossby number Ro, the average surface magnetic-field strength derived from the ZDI technique $\langle B_{\star}^{\rm ZDI} \rangle$, effective temperature $T_{\rm eff}$, stellar bolometric luminosity $L_\star$, inner $l_{\rm in}$ and outer $l_{\rm out}$ edges of the insulation HZ,  the upper limit of magnetospheric sizes $r_M $ (Eq.~(\ref{eq.rm})) of  planets with similar Earth's magnetisation orbiting in the HZ and the corresponding aperture of their auroral ovals $\alpha_0$. $R_{\rm orb}^{\min}$ is the closest orbital distance required for a terrestrial planet to sustain a present-day-Earth-sized magnetosphere (Eq.~(\ref{eq_rorbmin})). $B_{p,0}^{\min}$ and $B_{p,0}^{\min \dagger}$ are the range of the lowest planetary magnetic field  (Eq.~(\ref{eq.B_p})) required for a terrestrial planet orbiting inside the HZ to sustain the present-day ($11.7~r_p$) and young Earth's ($5~r_p$) magnetospheric size, respectively. 
\label{table_B}}    
\centering
\begin{tabular}{lllcccccccccccccc}  
\hline \hline
Star	& Year & Sp.&	$	M_* 	$	&	$	R_*		$ & $P_{\rm rot}$ & Ro& $\langle B_{\star}^{\rm ZDI} \rangle $	&$T_{\rm eff}$ &	$	\log{[{L_\star}/{L_\odot}]}		$			&	$	[l_{\rm in},	l_{\rm out}]$	& $	r_M^{\max}/r_p $ & $\alpha_0~(^\circ)$	&$R_{\rm orb}^{\min}$ & $B_{p,0}^{\min}$ (G)	& $B_{p,0}^{\min \dagger}$ (G)		\\		
ID		&obs.&type&	$(M_\odot)	$	&	$(R_\odot)	$&(d) &($0.01$)& (G)& (K)	&  &	(au)& $[B_{p,0}$=1G] & $[B_{p,0}$=1G]  &	(au)&	[present]&	[young]\\ \hline
    \object{GJ 182}  & $ 2007  $ & M0.5  & $ 0.75$ & $ 0.82$ & $ 4.35$ & $   17.4$ & $    172$ & $   3950$ & $-0.88$ &$[ 0.27   ,  0.73]$&$[  4.0   ,   7.7]$&$[   30   ,    21 ]$&$  1.3$ & $[   3.6    ,     25 ]$&$[  0.28    ,    2.0]$ \\
    \object{DT Vir}  & $ 2008  $ & M0.5  & $ 0.59$ & $ 0.53$ & $ 2.85$ & $    9.2$ & $    149$ & $   3790$ & $-1.26$ &$[ 0.18   ,  0.47]$&$[  5.5   ,    11]$&$[   25   ,    18 ]$&$ 0.48$ & $[   1.3    ,    9.6 ]$&$[  0.11    ,   0.75]$ \\
    \object{DS Leo}  & $ 2008  $ &   M0  & $ 0.58$ & $ 0.52$ & $14.00$ & $   43.8$ & $     87$ & $   3770$ & $-1.29$ &$[ 0.17   ,  0.46]$&$[  5.9   ,    11]$&$[   24   ,    17 ]$&$ 0.44$ & $[   1.1    ,    7.9 ]$&$[ 0.086    ,   0.62]$ \\
     \object{GJ 49}  & $ 2007  $ & M1.5  & $ 0.57$ & $ 0.51$ & $18.60$ & $   56.4$ & $     27$ & $   3750$ & $-1.31$ &$[ 0.17   ,  0.45]$&$[  6.1   ,    12]$&$[   24   ,    17 ]$&$ 0.42$ & $[  1.00    ,    7.2 ]$&$[ 0.078    ,   0.56]$ \\
    \object{OT Ser}  & $ 2008  $ & M1.5  & $ 0.55$ & $ 0.49$ & $ 3.40$ & $    9.7$ & $    123$ & $   3690$ & $-1.38$ &$[ 0.15   ,  0.41]$&$[  3.7   ,   7.1]$&$[   31   ,    22 ]$&$ 0.81$ & $[   4.4    ,     32 ]$&$[  0.35    ,    2.5]$ \\
    \object{CE Boo} & $ 2008  $ & M2.5  & $ 0.48$ & $ 0.43$ & $14.70$ & $   35.0$ & $    103$ & $   3570$ & $-1.56$ &$[ 0.13   ,  0.34]$&$[  3.0   ,   5.8]$&$[   35   ,    24 ]$&$ 0.90$ & $[   8.1    ,     59 ]$&$[  0.63    ,    4.6]$ \\
    \object{AD Leo}  & $ 2008  $ &   M3  & $ 0.42$ & $ 0.38$ & $ 2.24$ & $    4.7$ & $    180$ & $   3540$ & $-1.67$ &$[ 0.11   ,  0.30]$&$[  2.5   ,   4.9]$&$[   39   ,    27 ]$&$  1.0$ & $[    13    ,     97 ]$&$[   1.0    ,    7.6]$ \\
  \object{EQ Peg A}  & $ 2006  $ & M3.5  & $ 0.39$ & $ 0.35$ & $ 1.06$ & $    2.0$ & $    480$ & $   3530$ & $-1.73$ &$[ 0.10   ,  0.28]$&$[  2.2   ,   4.2]$&$[   43   ,    29 ]$&$  1.2$ & $[    21    ,    160 ]$&$[   1.7    ,     12]$ \\
    \object{EV Lac}  & $ 2007  $ & M3.5  & $ 0.32$ & $ 0.30$ & $ 4.37$ & $    6.8$ & $    490$ & $   3480$ & $-1.88$ &$[0.087   ,  0.24]$&$[  1.9   ,   3.8]$&$[   46   ,    31 ]$&$  1.2$ & $[    30    ,    220 ]$&$[   2.4    ,     17]$ \\
  \object{V374 Peg}  & $ 2006  $ &   M4  & $ 0.28$ & $ 0.28$ & $ 0.45$ & $    0.6$ & $    640$ & $   3410$ & $-2.02$ &$[0.074   ,  0.20]$&$[  1.7   ,   3.3]$&$[   50   ,    34 ]$&$  1.3$ & $[    45    ,    340 ]$&$[   3.5    ,     26]$ \\
  \object{EQ Peg B}  & $ 2006  $ & M4.5  & $ 0.25$ & $ 0.25$ & $ 0.40$ & $    0.5$ & $    450$ & $   3340$ & $-2.14$ &$[0.065   ,  0.18]$&$[  1.8   ,   3.4]$&$[   49   ,    33 ]$&$  1.0$ & $[    39    ,    290 ]$&$[   3.1    ,     23]$ \\
   \object{GJ 1156}  & $ 2009  $ &   M5  & $ 0.14$ & $ 0.16$ & $ 0.49$ & $    0.5$ & $    100$ & $   3160$ & $-2.60$ &$[0.038   ,  0.11]$&$[  3.3   ,   6.5]$&$[   33   ,    23 ]$&$ 0.23$ & $[   5.9    ,     45 ]$&$[  0.46    ,    3.5]$ \\
 \object{GJ 1245 B}  & $ 2008  $ & M5.5  & $ 0.12$ & $ 0.14$ & $ 0.71$ & $    0.7$ & $     60$ & $   3030$ & $-2.79$ &$[0.031   , 0.085]$&$[  3.5   ,   7.0]$&$[   32   ,    22 ]$&$ 0.17$ & $[   4.8    ,     37 ]$&$[  0.37    ,    2.9]$ \\
    \object{WX UMa}  & $ 2009  $ &   M6  & $ 0.10$ & $ 0.12$ & $ 0.78$ & $    0.8$ & $   1060$ & $   2770$ & $-3.10$ &$[0.022   , 0.061]$&$[  1.0   ,   1.7]$&$[   90   ,    50 ]$&$  1.0$ & $[   330    ,   2600 ]$&$[    25    ,    200]$ \\
    \object{DX Cnc}  & $ 2009  $ &   M6  & $ 0.10$ & $ 0.11$ & $ 0.46$ & $    0.5$ & $     80$ & $   2700$ & $-3.19$ &$[0.019   , 0.055]$&$[  2.8   ,   5.5]$&$[   37   ,    25 ]$&$ 0.16$ & $[   9.4    ,     75 ]$&$[  0.73    ,    5.9]$ \\
 \hline			
\end{tabular}
\end{sidewaystable}

\subsubsection{Stellar magnetic-field extrapolations} \label{sec.extrapolation}
We extrapolated the surface fields  into the stellar corona assuming the  stellar magnetic field $\mathbf{B}_\star$ is potential ($\nabla \times \mathbf{B}_\star =0$). By using this method, we ensured that the extrapolated magnetic fields are in their lowest-energy state, consistent with the conservative approach  described in Section \ref{sec.pl_mag}, in which we only considered the {\it lowest} external pressure that is exerted on the planet (Eq.~(\ref{eq.pextmin})). Stellar winds, whose dynamic pressure we neglect in the present work, are expected to stress the coronal magnetic fields, e.g., by the winding of the magnetic-field lines in the case of a rotating star. The wind, therefore, would remove the magnetic field from the lowest-energy state. 

Another effect caused by stellar winds is to stretch the magnetic-field lines in the radial direction. The general potential field extrapolation ignores this effect. As a way to overcome this deficiency, the potential field source surface (PFSS) method is often employed \citep[see, e.g.,][]{1969SoPh....9..131A, 1999MNRAS.305L..35J}. To emulate the stretching effect, beyond a radial distance $R_{\rm SS}$ (known as the source surface radius), the PFSS method assumes that the stellar magnetic field decays with distance squared.  At $R=R_{\rm SS}$, $B_{R,\star}$ is the only non-vanishing magnetic-field component and magnetic-flux conservation implies that for, $R_{\rm orb}>R_{\rm SS}$, the magnetic field of the stellar corona is given by
\begin{equation}\label{eq.BbeyondSS}
B_\star(R_{\rm orb}>R_{\rm ss}, \theta, \varphi) = B_{R,\star} (R_{\rm SS}, \theta, \varphi) \left( \frac{R_{\rm SS}}{R_{\rm orb}}\right)^2.
\end{equation}
We note that $B_{R,\star} (R_{\rm SS}, \theta, \varphi)$ is a function not only of the distance to the source surface, but also of the co-latitude $\theta$ and longitude $\varphi$. For our estimates from now on, {we assume an average magnetic-field strength at $R_{\rm SS}$ and, for shortness, we write $|B_{\rm SS}| =  \langle |B_{R,\star} (R_{\rm SS}, \theta, \varphi)|\rangle$}. The magnetic pressure is thus 
\begin{equation}\label{eq.pb_pfss}
P_{B,\star} (R_{\rm orb})= \frac{B_{\rm SS}^2}{8\pi} \left( \frac{R_{\rm SS}}{R_{\rm orb}}\right)^4. 
\end{equation}

\begin{figure}
\centering
\includegraphics[width=\hsize]{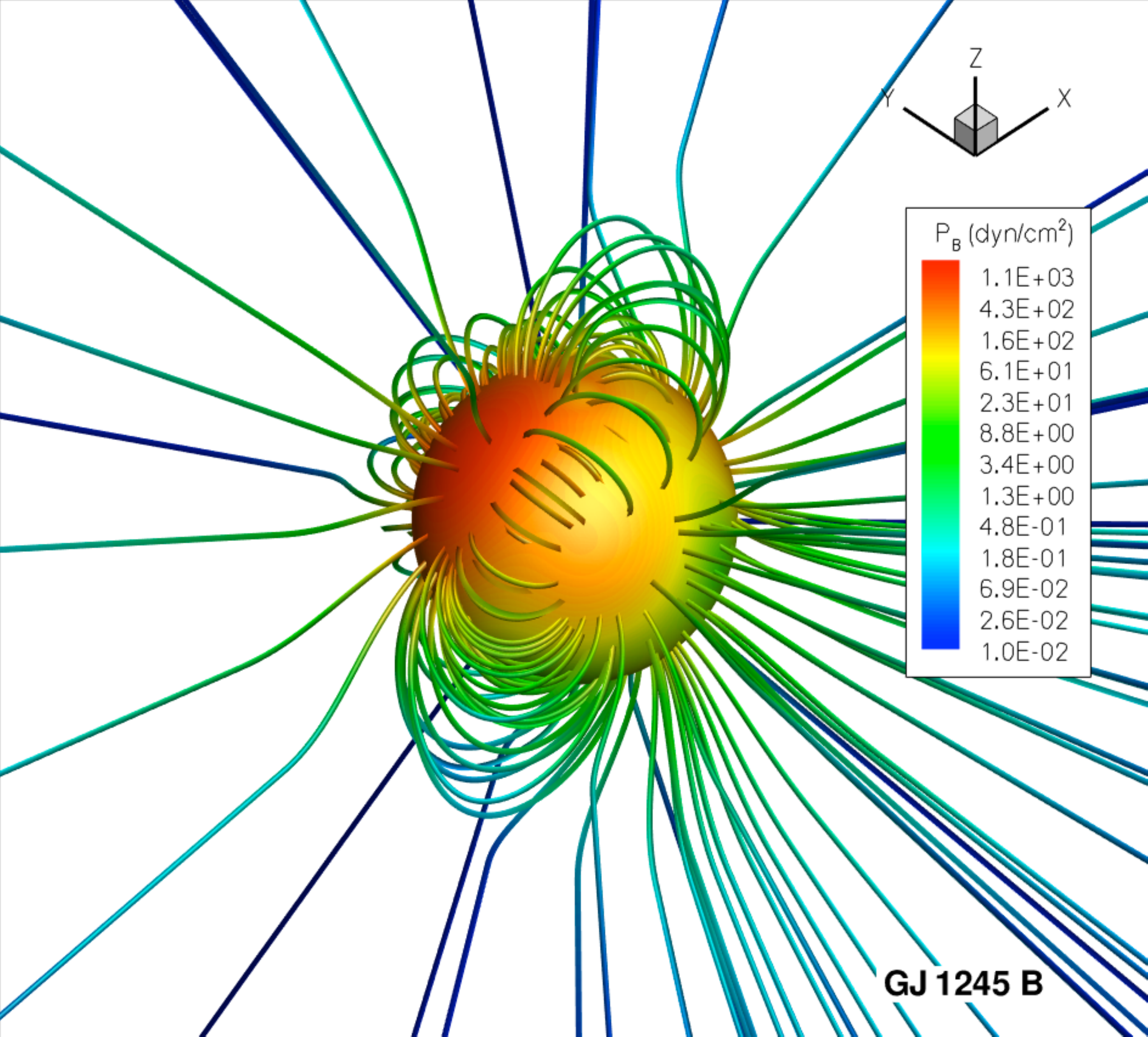}
  \caption{Stellar coronal magnetic field lines of GJ~1245~B extrapolated using the potential field source surface method. The colour is coded according to the local value of the magnetic pressure that would be exerted on a planet orbiting this star.
\label{fig.magnetograms}}
\end{figure}

Throughout the present work, we assume $R_{\rm SS}=2.5~R_\star$, and we note that for all  objects investigated here, $R_{\rm orb}$ is sufficiently large such that $R_{\rm orb}>R_{\rm SS}$. In Appendix~\ref{appendix}, we show that a physically reasonable different choice of $R_{\rm SS}$ does not change our conclusions. Figure~\ref{fig.magnetograms} shows the magnetic-field lines of the late-dM star GJ~1245~B as an illustration of the extrapolation method. Values of the local  $P_{B,\star}$ are colour-coded. For the early- and mid-M stars considered in this paper, the visualisation of potential field extrapolations into the stellar corona can be found in \citet{2012MNRAS.424.1077L}. 

\section{Active M-dwarf planet hosts}\label{sec.results}
\subsection{Magnetospheric characteristics}\label{subsec.part1}

Eq.~(\ref{eq.rm}) provides a useful upper limit on the size of magnetospheres of planets orbiting the dM stars in our sample. Because we still have little information about magnetic fields of terrestrial planets orbiting dM stars, we took the Earth as an example and assumed that the hypothetical planets orbiting the stars in our sample have magnetic fields similar to the Earth ($B_{p,0}\sim1~G$). 

\begin{figure}
\centering
\includegraphics[width=\hsize]{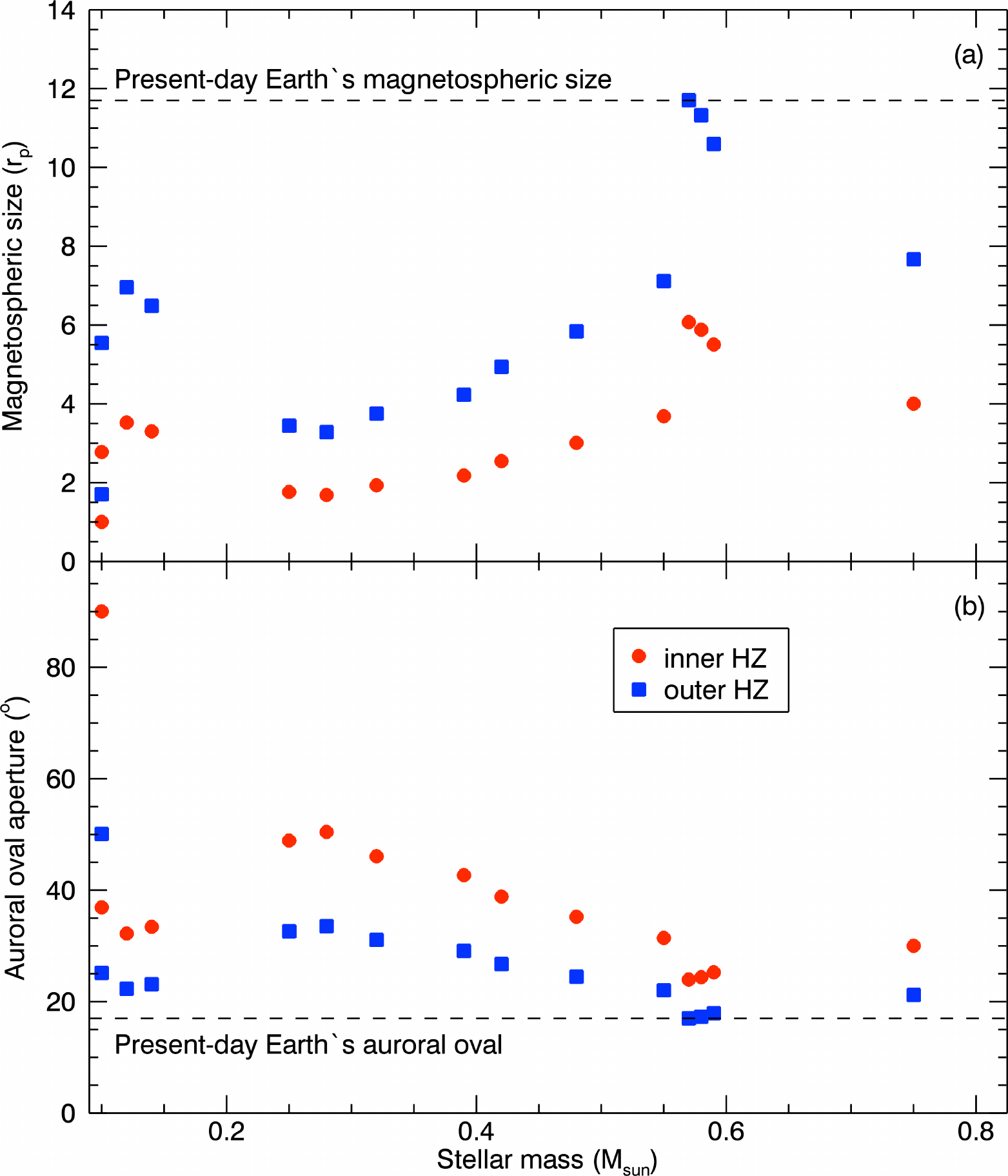}
\caption{(a) Upper limit of magnetospheric sizes and (b) the corresponding apertures of their auroral ovals for hypothetical planets orbiting the dM stars in our sample. We assume that these planets harbour dipolar magnetic fields with the same intensity as that of the Earth. For comparison, the dashed line shows the present-day Earth's magnetospheric size (upper panel) and the Earth's auroral oval aperture (lower panel).\label{fig.rM_size}}
\end{figure}

We calculated the extent of the HZ following the prescription of \citet{2007A&A...476.1373S} based on the early Mars and recent Venus criteria. Table~\ref{table_B} presents the inner ($l_{\rm in}$) and outer $(l_{\rm out})$ edges of the HZ for the stars in our sample, using the listed stellar luminosities and effective temperatures. Figure~\ref{fig.rM_size}a shows the upper limit of $r_M$ if the Earth-analogue planets were orbiting the inner (circles) and outer (squares) edges of the HZ. The present-day magnetospheric size of the Earth, taken to be $r_M =  11.7~r_p$ throughout the present paper, is marked by the dashed line. We note that almost all the hypothetical planets would have magnetospheric sizes considerably smaller than that of the Earth: in the limit they were orbiting at the inner edge of the HZ, their magnetospheric sizes would extend at most up to $6.1~r_p$, while planets orbiting at the outer edge would present a maximum magnetospheric size of $11.7~r_p$ (Table~\ref{table_B}). 
 
In addition to the extent of the magnetosphere of the planet, we also calculated the amount of planetary area that remains unprotected because the planetary magnetic-field lines stay open, allowing for particles to be transported to/from the interplanetary space. Similar calculations have been presented in \citet{1975JGR....80.4675S}, \citet{2010Sci...327.1238T}, \citet{2011MNRAS.414.1573V}, and \citet{2013arXiv1304.2909Z}. {As in those works, here we took the colatitude of the open-closed field line boundary to be the colatitude of the auroral oval ring (first-order approximation), but we recall that these two colatitudes may not  match exactly \citep[e.g.][]{1975JGR....80.4675S, 2001JGR...106.8101H}.} The aperture of the auroral ring can be related to the size of the planet magnetosphere $r_M$ as
\begin{equation}\label{eq.alpha}
\alpha_0 = \arcsin \left[ \left( \frac{r_p}{r_M} \right)^{1/2} \right] \, ,
\end{equation}
which implies in a fractional area of the planetary surface that has open magnetic field lines 
\begin{equation}\label{eq.area}
\frac{A_{\rm polar}}{A_{\rm planet}} = (1-\cos \alpha_0),
\end{equation} 
 where we considered both the northern and southern auroral caps of the planet. We note that a stronger external pressure not only makes $r_M$ smaller, but also exposes a larger area of the polar cap of the planet. For the Earth, the aperture of the auroral oval is $\alpha_0 \simeq 17$ -- $20^{\rm o}$ \citep{2009AnGeo..27.2913M}, which implies that the open-field-line region covers only $\sim 5$ -- $6 \%$ of the surface. Figure~\ref{fig.rM_size}b shows the aperture of the auroral oval that the hypothetical planets would present if they were orbiting the inner (circles) and outer (squares) edges of the HZ. For that, we again assumed a magnetisation similar to the terrestrial magnetisation.  The present-day auroral oval of the Earth $\alpha_0 \simeq 17 ^{\rm o}$ is marked by the dashed line. The calculated apertures (Table~\ref{table_B}) range from $24^\circ$ to $90^\circ$ for planets at the inner boundary of the HZ, which can be significantly larger than the Earth's auroral oval aperture. These limits correspond to fractional open areas ranging from $9\%$ up to $100\%$ -- the latter case corresponds to a scenario where the intrinsic planetary magnetosphere is crushed into the planet surface. For planets orbiting in the outer edge of the HZ, $\alpha_0$ ranges from $17^\circ$ to $50^\circ$ and the corresponding fractional area of the planet with open field lines ranges from $4\%$ to $36\%$. 

Our calculations neglect the effects of the stellar wind ram pressure and, therefore, provide upper (lower) limits of the magnetospheric sizes (auroral oval apertures). A crude estimate of the effects that the consideration of the stellar wind ram pressure would have in our results is presented next. Taking a typical M4.0 star with an isothermal stellar wind ($10^6$~K), \citet{2011AN....332.1055V} estimated a stellar wind velocity of $\sim 380$~km~s$^{-1}$ and a density $4.5\times 10^{-7}~n_0$ at the inner edge of the HZ, where $n_0$ is the base density of the stellar wind. Because densities (and mass-loss rates) are not clearly constrained in dM stars \citep[see the discussion in][]{2011MNRAS.412..351V}, we assumed wind base density values that are two orders of magnitude lower and three orders of magnitude higher than the solar wind base density $n_{0,\odot} \simeq 10^8$~cm$^{-3}$ \citep{1988ApJ...325..442W}. We find the ram pressure of the stellar wind to be roughly in the range $\sim 5\times10^{-16} n_{0}  \sim 5\times 10^{-10} -5\times 10^{-5}$~dyn~cm$^{-2}$ at the inner edge of the HZ. For the stars in our sample, the average magnetic pressure at the inner edge of the HZ is in the range $2\times10^{-7}$ -- $3\times 10^{-2}$~dyn~cm$^{-2}$. Combined, the stellar ram and magnetic pressures would act to reduce the sizes of the magnetospheres derived here by up to a factor of $2.5$ and to increase the size of the auroral oval aperture by up to a factor of $1.7$. The largest effects are noticed in the least magnetised cases. We recall that although these estimates illustrate the effects of the stellar wind ram pressure on the magnetospheric sizes, they are very crude -- to properly quantify these effects, a more complete wind modelling that includes rotation and magnetic effects is desirable \citep{2011MNRAS.412..351V}. 

Although we only calculated upper limits of the magnetospheric sizes and the corresponding lower limits of auroral oval apertures, our results show that Earth-like planets with similar terrestrial magnetisation orbiting active dM stars present {\it smaller} magnetospheric sizes than that of the Earth. Section \ref{sec.cond} investigates the conditions required for planets orbiting the stars in our sample to present Earth-sized magnetospheres.

\subsection{Conditions for Earth-sized magnetospheres}\label{sec.cond}
\subsubsection{Orbital distances}
\begin{figure}
\centering
\includegraphics[width=\hsize]{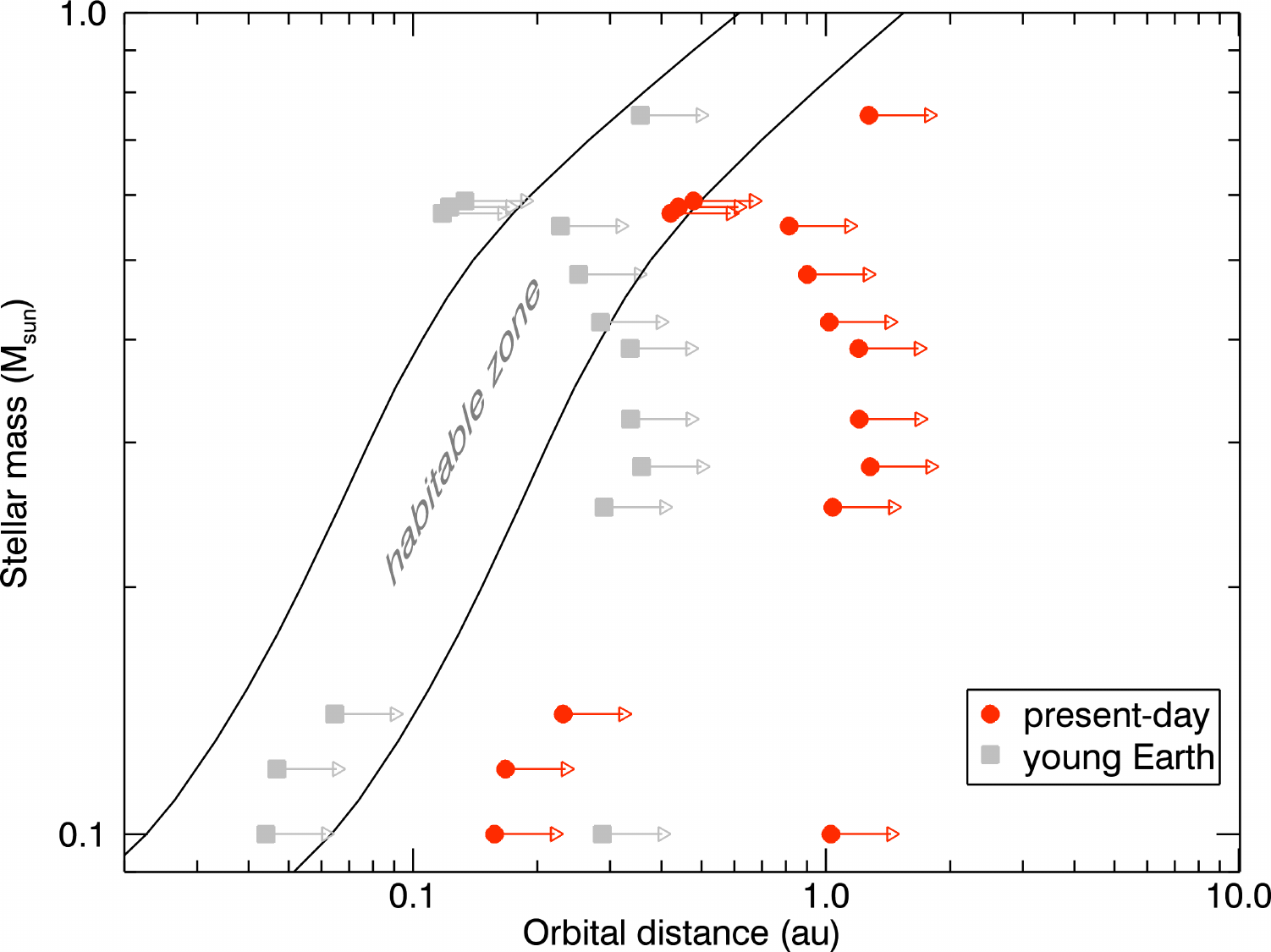}
\caption{Closest orbital distance at which an Earth-like planet orbiting the stars in our sample would be able to sustain the present-day (red circles)  and the young (3.4~Gyr ago, grey squares) Earth's magnetospheric size, assuming it has the same magnetic field as the Earth. 
We use a sample of 15 active stars for which large-scale surface magnetic-field maps were reconstructed, to determine the magnetic pressure at the planet orbit and hence the largest extent of its magnetosphere. Planets orbiting at a closer orbital radius would experience a stronger stellar magnetic pressure, which could reduce the size of the planet's magnetosphere significantly, exposing the planet's atmosphere to erosion by the stellar wind. For reference, we show the inner/outer edge of the HZ for $1$-Gyr-old low-mass stars (solid lines). \label{fig.HZ}}
\end{figure}

In the Earth's case, the ram pressure of the solar wind dominates the external pressure contribution in defining the size of the magnetosphere. The ram pressure required to sustain the Earth's magnetosphere ($r_M=11.7~r_p$ and $B_{p,0}\sim 1$~G) is $P_{{\rm ram, }\odot} \sim 3.9 \times 10^{-9}$~dyn~cm$^{-2}$. However, for planets orbiting in the HZ of the dM stars in our sample, we showed in Section~\ref{subsec.part1} that the high stellar magnetic pressure alone is sufficient to cause a greater reduction in the size of the magnetosphere. Therefore, planets orbiting within the radius where $P_{B,\star} (R_{\rm orb})  = P_{{\rm ram, }\odot}$ would have magnetospheres that are smaller than the current magnetospheric size of the Earth. We use Eq.~(\ref{eq.pb_pfss}) to derive the {\it minimum} orbital distance beyond which $P_{B,\star} < P_{{\rm ram, }\odot}$ 
\begin{equation} \label{eq_rorbmin}
\frac{R_{\rm orb}^{\min}}{R_\star} = \left( \frac{ B_{\rm SS}^2}{8\pi P_{{\rm ram, }\odot}} \right)^{1/4}\left(\frac{R_{\rm SS}}{R_\star}\right) .
\end{equation}
The red circles in Figure~\ref{fig.HZ} mark the position beyond which an Earth-like planet orbiting the stars in our sample would be able to sustain an Earth-sized magnetosphere, assuming it has the same magnetic field as the Earth (values of ${R_{\rm orb}^{\min}}$ are also shown in Table~\ref{table_B}). For reference, the solid lines in Figure~\ref{fig.HZ} delimit the extent of the insulation HZ following the prescription of \citet{2007A&A...476.1373S}, coupled to the stellar evolution models from \citet{1998A&A...337..403B} for $\sim 1$~Gyr-old stars.  We note that with the exception of only a few cases, the terrestrial planet orbiting the stars in our sample would need to orbit significantly farther out than the traditional limits of the HZ to be able to sustain an Earth-sized magnetosphere.

\subsubsection{Planetary magnetic fields}\label{sec.results_b}
If these hypothetical planets were orbiting within the HZ, where the stellar magnetic pressure is higher, the sizes of their magnetospheres would be smaller, which could expose the planet's atmosphere to erosion by the stellar wind. To counterbalance this higher external pressure of the HZ, a terrestrial planet requires a higher magnetic field to present an Earth-sized magnetosphere. From Eq.~(\ref{eq.rm}), we derive the {\it lowest} required intensity of the planetary magnetic field of an Earth-sized magnetosphere
\begin{equation}\label{eq.B_p}
\frac{B_{p,0}^{\min}}{[1~{\rm G}]} \simeq 16052 \left[\frac{P_{B,\star} (R_{\rm orb})}{{\rm [1~dyn~cm}^{-2}]} \right]^{\frac12},
\end{equation}
where $P_{B,\star}$ is given by Eq.~(\ref{eq.pb_pfss}). Table~\ref{table_B}  shows the range of $B_{p,0}^{\min}$ required to balance the external pressure exerted by the stellar magnetic field {\it alone}, assuming an orbital radius located between $l_{\rm in}$ (highest value in the presented range) and $l_{\rm out}$ (lowest value). A planet orbiting at $R_{\rm orb} = l_{\rm in}$, which is closer to the stars in our sample by a factor of $2.65$ -- $2.82$ than for an orbit at $R_{\rm orb} = l_{\rm out}$, requires a magnetic field that is stronger by a factor of $7$ -- $8$. We find that the required $B_{p,0}^{\min}$ for these hypothetical planets can be significantly larger than the Earth's, depending on their orbital radius and the host-star magnetism. Although the magnetic field is a quantity that is still poorly known for exoplanets (see Section \ref{sec.intro}), theoretical work suggests that super-Earths are not expected to host such strong dipolar fields \citep{2012Icar..217...88Z}. 
 
From Eqs.~(\ref{eq.pb_pfss}) and (\ref{eq.B_p}), we note that $B_{p,0}^{\min}$ decays with the squared normalised orbital distance in stellar radii and, for a dipolar field, it increases with the cube of the normalised size of the magnetosphere in planetary radii. Therefore, $B_{p,0}^{\min}$ is more sensitive to variations on $r_M/r_p$ than on $R_{\rm orb}/R_\star$. 

\subsection{Conditions for young-Earth-sized magnetospheres}\label{sec.polarcap} 
In our estimates up to here, we have assumed that the fictitious Earth-like planet orbiting the dM stars in our sample has a magnetosphere extending out to $11.7~r_p$ (similar to the Earth). However, \citet{2007AsBio...7..185L} suggested that magnetospheric sizes of $\gtrsim 2~r_p$ can already offer a reasonable protection for the planetary atmosphere. If this is indeed the case, $B_{p,0}^{\min}$ derived in Section \ref{sec.results_b} would decrease by a factor $(11.7/2)^3\simeq 200$. In this case, even for a planet orbiting in the inner edge of the HZ, the derived values for minimum planetary magnetic fields would be much closer to the magnetic field intensities found in the planets in the solar system. 

If we consider that $r_M\simeq 2~r_p$ can still offer a reasonable protection for the planetary atmosphere, as suggested by \citet{2007AsBio...7..185L}, then such a planet would present an auroral oval aperture of $\alpha_0\simeq 45^{\rm o}$ and the open field line region would cover $\sim 30 \%$ of the planetary surface: a significantly larger area of the planet would remain exposed to, e.g., incidence of particles from the star (generated in flares, CMEs, stellar wind) and from the cosmos (galactic cosmic rays), as well as escape of planetary atmosphere through polar flows \citep{2005AsBio...5..587G,2009Icar..199..526G,2007RvGeo..45.3002M,2007AsBio...7..167K,2007AsBio...7..185L}.
 
The reduced size of magnetospheres and consequent increase of the polar cap area are believed to be accompanied by an increase of the volatile losses from the exosphere, which can affect long-term atmospheric composition \citep{2010Sci...327.1238T}.  At this point, the smallest area of the planetary surface that could be exposed (and for how long it is allowed to last) before it starts affecting the potential for formation and development of life in a planet is unknown.  An area of $\sim 30 \%$  of exposed surface might have effects on life on the planet, but it is beyond the scope of the present paper to assess this effect. 

\citet{2010Sci...327.1238T} claimed that $3.4$~Gyr ago, the magnetospheric size of the young Earth was smaller than the present-day value, possibly extending out to $\sim 5~r_p$. 
Based on the hypothesis that such a reduced magnetospheric size would still be suitable to prevent escape of a significant amount of volatile content from the planetary exosphere, as well as generating a sufficiently small auroral cap, we recalculated $B_{p,0}^{\min}$ for planets orbiting in the HZ of the stars in our sample, assuming a young-Earth magnetospheric size. We find that the auroral oval extends to colatitudes of about $27^{\rm o}$, corresponding to a polar cap area of $11\%$ (Eqs.~(\ref{eq.alpha}) and (\ref{eq.area})). In that case, the lowest planetary magnetic field ($B_{p,0}^{\min \dagger}$) required to sustain such a magnetosphere (last column of Table~\ref{table_B}) is a factor of $13$ lower than the values derived in Section \ref{sec.results_b}, where the present-day size of Earth's magnetosphere was assumed instead. 

The grey squares in Figure~\ref{fig.HZ} show the shortest orbital distance at which a terrestrial planet orbiting the stars in our sample would be able to sustain a young-Earth magnetospheric size. We find that about $2/3$  of the cases now lie inside or near the outer edge the HZ. We recall, however, that because our approach only considers a lower limit for the external pressure, these points are likely to move outward because of stellar wind ram and thermal pressures.

\citet{2010Sci...327.1238T}  argued further that the reduced size of the Earth's magnetosphere would have remained that way on a time scale of millions to tens of millions of years. Although such a reduced magnetosphere did not prevent formation and development of life in Earth, we do not know what the effects would have been if it persisted for longer periods of time (e.g., a few Gyr).
 
\section{Weakly-active M-dwarf planet hosts}\label{sec.results2}
The sample of stars considered in Section \ref{sec.results} mostly includes rapidly rotating stars that are in general more active -- because these stars are the most accessible to ZDI studies. It is expected that as a star ages, its wind removes stellar angular momentum, spinning the star down. Below a critical value of the rotation rate, the dynamo is expected to be less efficient, resulting in weaker magnetic fields, which decay along with rotation rates. However, the time scale for that to happen seems to be very long, as dM stars are observed to remain active (and therefore, rapidly rotating) for a long duration of time. For example, stars with spectral types M5 -- M7 are believed to remain active/rapidly rotating for $\sim 6$ to $10$ Gyr \citep{2008AJ....135..785W,2011ApJ...727...56I}. 

\citet{2009ApJ...692..538R} showed that rapidly rotating dM stars (spectral types in the range M0 to M7) with Rossby numbers ${\rm Ro}<0.1$ have an average surface magnetic field strength $\langle B_{\star} \rangle$ that remains independent of rotation rate. For these stars, $ \langle B_{\star} \rangle = \langle B_{\rm crit} \rangle \sim 3$~kG, with a scatter of about 1 kG. Note that, with the exception of a few early dM stars, all stars considered in Section~\ref{sec.results} are in the so-called saturated activity regime. We can relate the average surface magnetic field strength with the rotation period $P_{\rm rot}$ of the star as \citep{2012ApJ...746...43R}
\begin{eqnarray}\label{eq.crit}
\langle B_\star \rangle = \langle B_{\rm crit} \rangle, ~{\rm for}~P_{\rm rot} \le P_{\rm rot}^{\rm (crit)}   \nonumber \\ 
\langle B_\star \rangle = \langle B_{\rm crit} \rangle \left( \frac{P_{\rm rot}^{\rm (crit)}}{P_{\rm rot} } \right)^a, ~{\rm for}~P_{\rm rot}  > P_{\rm rot}^{\rm (crit)} \, ,
\end{eqnarray}
where we adopt $P_{\rm rot}^{\rm (crit)} \simeq 10$~days, $a\sim 1.7$, and $ \langle B_{\rm crit} \rangle \sim 3$~kG, based on empirical findings from \citet{1996IAUS..176..237S}, \citet{2003A&A...397..147P}, and \citet{2009ApJ...692..538R}. We note here that saturation is a function of stellar mass and that it likely starts at longer $P_{\rm rot}^{\rm (crit)}$ for the lowest mass range \citep{2007AcA....57..149K}. For the {ultracool dwarfs}, i.e., dM stars with spectral types later than M7, magnetic-field generation seems to be untied to rotation \citep{2010ApJ...710..924R}. These stars were excluded from our study.

We note that $\langle B_\star \rangle$ refers to the average magnetic fields derived from unpolarised spectroscopy (e.g., absorption lines of molecular FeH), a technique that is capable of measuring the unsigned magnetic flux (including the dominant fraction at small scales), but is mostly insensitive to the field topology. The ZDI technique, on the other hand, probes the topology of the large-scale surface field $\langle B_{\star}^{\rm ZDI} \rangle $. At the position of the planet, this is the component of the stellar magnetic field that still survives, because the small-scale structure rapidly decays with height (Lang et al. 2013, in prep.). To relate $\langle B_{\star}^{\rm ZDI} \rangle $  with the average small-scale field $\langle B_\star \rangle$ from Eq.~(\ref{eq.crit}), we used the empirically determined ratio $f=\langle B_{\star}^{\rm ZDI} \rangle  / \langle B_\star \rangle \sim 6\%$ \citep[Figure~16 in][]{2010MNRAS.407.2269M}. Although this ratio  was derived from a handful of active stars, it appears to be valid in the unsaturated regime (or at least in its upper part). Note that higher values of $f\sim15$ -- $30\%$ were found for fully convective stars with dipole-dominated large-scale fields, but it is not clear whether ratios as high as this exist among the weakly active stars discussed in this Section. 

Our procedure was the following. We estimated the external magnetic field strength that provides for a planet in the HZ of a dM star the present-day value of the external pressure at the Earth ($P_{\rm ext} = P_{{\rm ram,}\odot}$).  To estimate the large-scale surface stellar magnetic field, we extrapolated this local field back to the stellar surface using the PFSS method. In this section, we assume for simplicity that the stellar large-scale surface magnetic field is dipolar and adopted $R_{\rm SS}=2.5~R_\star$. Once the large-scale surface field was determined, we used the empirical ratio between $\langle B_{\star}^{\rm ZDI} \rangle$ and $ \langle B_\star \rangle$ of $f\sim6\%$ and the rotation-magnetic field relation (Eq.~(\ref{eq.crit})) to estimate the stellar rotation period. This is the shortest period (longest rotation rate) for which a large-scale dipolar field would provide for a planet in the HZ of a dM star the same external pressure as the present-day value of the external pressure at the Earth. In this case, a planet with a similar magnetisation as that of the Earth would present an Earth-sized magnetosphere.

\begin{figure}
\centering
\includegraphics[width=\hsize]{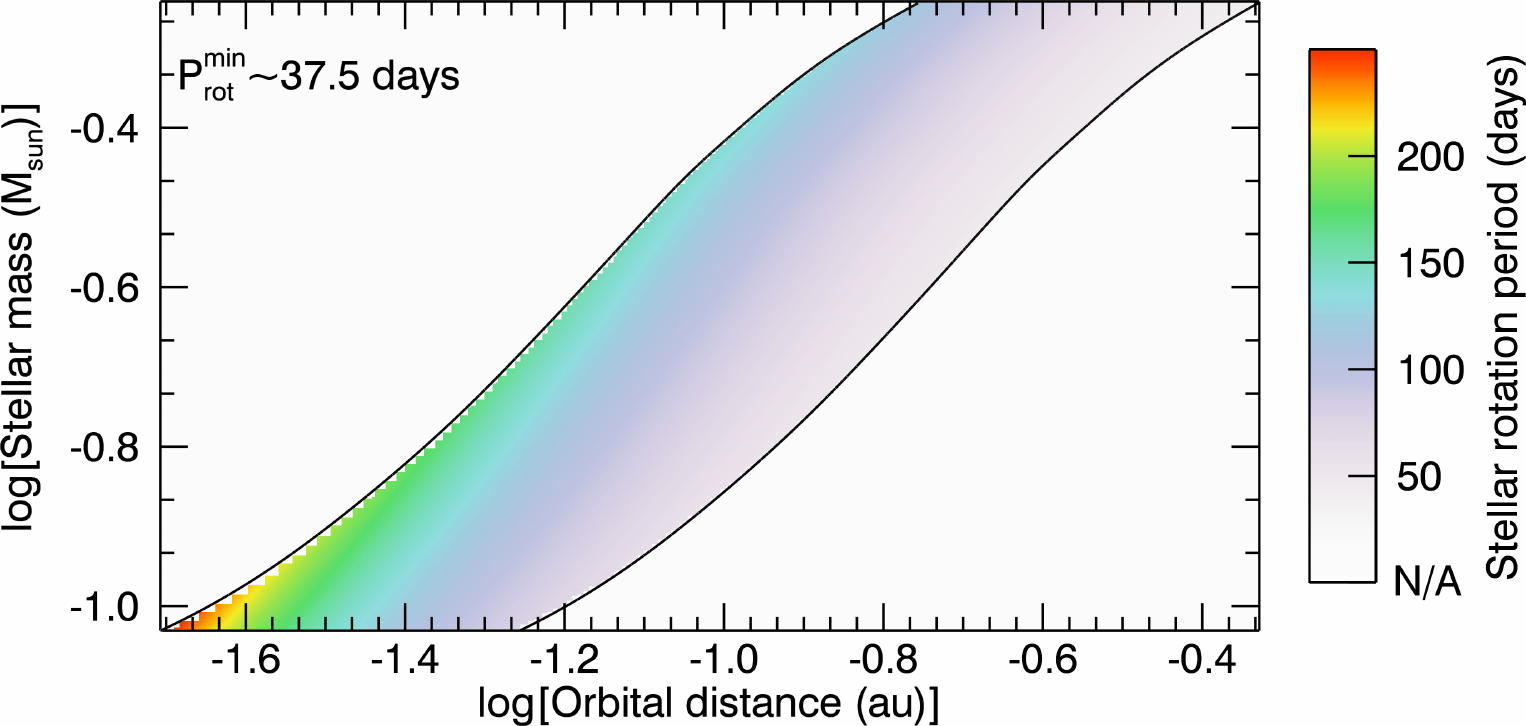}
\caption{Solid lines: edges of the HZ for low-mass stars (spectral types earlier than M7). Inside this region, we show the shortest stellar rotation period for which the stellar magnetic field pressure causes the planetary magnetosphere to have the same size as the Earth's ($B_{p,0}\sim 1$~G). Stellar periods must be longer than about $37$ -- $202$~days ($\gtrsim 62$ -- $263$~days) for early- and mid-dM stars (late-dM stars). Because the fraction of inactive (i.e., slowly rotating) dM stars decays considerably for late-dM stars, conditions for terrestrial planets to harbour Earth-sized magnetospheres are more easily achieved for planets orbiting slowly rotating early- and mid-dM stars.
\label{fig.HZ_prot}}
\end{figure}

Figure \ref{fig.HZ_prot} shows how the required stellar rotation period varies as a function of stellar mass and orbital distance for planets in the HZ. To derive the stellar radius, mass, temperature and luminosity, we adopted the evolutionary models of \citet{1998A&A...337..403B} coupled to the habitability prescription by \citet{2007A&A...476.1373S}. Interestingly, we note that the required rotation rate of the early- and mid-dM stars ($\gtrsim 37$ -- $202$~days, depending on $R_{\rm orb}$) is slower than the solar rate, while for the late-dM stars, the required stellar rotation rate can be even slower ($\gtrsim 62$ -- $263$~days). Planets with Earth-like magnetic fields orbiting dM stars that rotate faster than these values  would have smaller magnetospheric sizes than that of the Earth. 

Note that the rotation period of late-dM stars would be even longer if the mass-dependence of $P_{\rm rot}^{\rm (crit)}$ \citep[or, equivalently, the convective turnover time,][]{2007AcA....57..149K} were considered. Although there is a number of known old, slowly rotating late-dM stars with rotation periods of up to $100$~days \citep{2011ApJ...727...56I, 2012MNRAS.427.3358G, 2013MNRAS.tmp.1203M}, the fraction of inactive dM stars is considerably lower for late-dM stars (\citealt{2008AJ....135..785W}). Furthermore, because their rotational braking time scales are probably longer \citep[$\sim 6 $ -- $10$~Gyr, ][]{2008AJ....135..785W,2011ApJ...727...56I} than the lifetime of the dipole-dominated planetary fields \citep[$\lesssim 3$~Gyr,][]{2012Icar..217...88Z}, it may be more difficult for an old ($\gtrsim 6$~Gyr) Earth-like planet to generate a significant dipolar magnetic field.
Therefore, our results indicate that conditions for terrestrial planets to harbour Earth-sized magnetospheres (and larger) are more easily achieved for planets orbiting the early- and mid-dM stars.

It is useful to express our results in an analytical form, which can be used to estimate the shortest stellar rotation period for which the large-scale stellar field {is taken to be the only} external pressure at the orbit of the planet. Analytically, we have  
\begin{equation}
P_{\rm rot}^{\min} \simeq \left[ \left(\frac{f \langle B_{\rm crit}\rangle}{\sqrt{8 \pi}}\right)^{1/a} P_{\rm rot}^{\rm (crit)}\right]  \left[ \frac{\langle B_{\star}^{\rm ZDI} (R_{\rm SS})\rangle }{\langle B_{\star}^{\rm ZDI} \rangle} \frac{R_{\rm SS}^2}{\sqrt{P_{{\rm ext}}}}\right]^{1/a}\frac{1}{{R_{\rm orb}}^{2/a}},
\end{equation}
where we grouped observable quantities (term inside first brackets) and quantities that are model-dependent (term inside second brackets). The previous equation reduces to
\begin{equation}\label{eq.analytic-prot}
P_{\rm rot}^{\min} \simeq  {\mathcal{C}}{(R_\star/R_{\rm orb})^{2/a}},
\end{equation}
where $\mathcal{C}$ is a coefficient dependent on the large-scale field topology, the choice of $R_{\rm SS}$ and the external pressure.  For a dipolar field with $R_{\rm SS}=2.5~R_\star$ and the same external pressure as the present-day value of the external pressure at the Earth, $\mathcal{C}\simeq 17800$~days, while for an external pressure similar to that of the young Earth,  $\mathcal{C}\simeq 4000$~days.

Eq.~(\ref{eq.analytic-prot}) and Figure \ref{fig.HZ_prot} show that for a star of given mass, the stellar rotation period required to allow an Earth-sized magnetosphere is longer for a planet orbiting in the inner edge of the HZ than it would be if the planet were orbiting farther out. Additionally, for planets orbiting at the same physical distances (in au), the required stellar rotation period decreases for lower-mass stars, i.e., a slowly rotating early-dM star (larger masses) can produce the same external pressure for a terrestrial planet as a more rapidly rotating late-dM star (lower masses). This is a consequence of our model, which only considers the effects of the stellar magnetic pressure: the stellar magnetic-field intensity does not require the physical distance, but is rather a function of the normalised distance with respect to the stellar radius (see, e.g., Eqs.~(\ref{eq.BbeyondSS}), (\ref{eq.pb_pfss}) and (\ref{eq.analytic-prot})). Therefore, if two stars have the same large-scale magnetic-field strength and, therefore, the same rotation period (Eq.~(\ref{eq.crit})), but have different radii, a planet orbiting at the same physical distance will experience a less intense magnetic field if it were orbiting the smaller star. 

It is interesting to note that the Neptune-sized planet GJ~674b orbits the most active planet-host dM star (M2.5, $P_{\rm rot} \simeq 35$~days) at a distance $R_{\rm orb} \simeq 0.039\simeq 25~R_\star$  \citep{2007A&A...474..293B}. Eq.~(\ref{eq.analytic-prot}) shows that the shortest required rotation period of GJ~674 is  $P_{\rm rot}^{\min}\simeq 395$~days for GJ~674b to have a present-day Earth-sized magnetosphere and $P_{\rm rot}^{\min}\simeq 88$~days for a young-Earth-sized one (assuming $B_p\simeq 1$~G). Because $P_{\rm rot}^{\min}\gtrsim P_{\rm rot}$,  GJ~674b should present a magnetosphere that is smaller than the young-Earth's one. The somewhat large estimated $P_{\rm rot}^{\min}$ is due to the close planetary orbit, which lies closer than the HZ limits (ranging from $\sim 60$ to $160~R_\star$). An additional planet orbiting within the HZ of GJ~674 could harbour young-Earth-sized magnetosphere ($ P_{{\rm rot}} \gtrsim P_{\rm rot}^{\min} \simeq 10$ -- $33$~days), but our model predicts that it would be more difficult for it to harbour a present-day-Earth-sized magnetosphere (because $ P_{{\rm rot}} \lesssim P_{\rm rot}^{\min} \simeq 45$ -- $146$~days).

\section{Summary and final remarks}\label{sec.conc}
We quantitatively evaluated the extent of planetary magnetospheres as a result of the pressure of intense stellar magnetic fields found around M dwarf (dM) stars. {If a planet possesses a sufficiently strong magnetic field, it will have a magnetosphere, one role of which is deflecting coronal material around the planet. Note that even within the solar system, the details of magnetospheric systems vary significantly, and a complete analysis should consider current systems, planet rotation, bow shocks, and plasma sources such as moons or the planet itself, and should properly account for stellar wind forces. At present, little is known about many of these features for exoplanets and we did not speculate about them here. However, a reasonable first-order estimate of their magnetospheric size may be obtained using a much simpler pressure-balance calculation, providing a fast means to assess which planets are most affected by the stellar magnetic field without requiring sophisticated simulations.} Our main findings are summarised as follows:

\begin{enumerate}
\item In Section \ref{sec.results}, we investigated the magnetic environment of 15 active dM stars (spectral types M0 to M6), for which surface magnetic-field maps were observationally reconstructed in the literature. We showed that a planet with a similar magnetic-field intensity as that of the Earth that were orbiting any of these stars would present a magnetosphere that would extend at most up to $6.1~r_p$ ($11.7~r_p$) if it were orbiting at the inner (outer) edge of the habitable zone (HZ). The corresponding sizes of the auroral oval  ranges from $24^\circ$ to $90^\circ$ ($17^\circ$ to $50^\circ$), exposing from about $9\%$ up to $100\%$ ($4\%$ to $36\%$) of the planetary area to, e.g., incidence of particles from the star and from the cosmos as well as escape of planetary atmosphere through polar flows.

\item With the exception of only a few cases, we showed that if a terrestrial planet were orbiting one of these stars, the Earth-like planet would need to be orbiting much beyond the insulation HZ (Figure~\ref{fig.HZ})  for it to have an Earth-analogue magnetosphere (i.e., same terrestrial magnetic-field strength and magnetospheric size). 

\item We also showed that if the Earth-like planet were required to orbit {\it inside} the HZs of these stars, it would need a significantly stronger magnetic field to reach the size of the present-day Earth magnetosphere ($r_M \simeq 11.7~r_p$). The derived magnetic fields range from a few G to up to a few thousand G (see Table~\ref{table_B}).

\item The young Earth is believed to have had a magnetospheric size that is smaller than its current value ($r_M \simeq 5~ r_p$). In this case, the polar-cap area of the planet that is unprotected from transport of particles to/from interplanetary space would be twice as large. By adopting a condition more suitable for a young-Earth analogue magnetosphere, we showed that terrestrial planets orbiting inside the HZ of the stars in our sample would require planetary magnetic fields one order of magnitude smaller than the ones found considering the present-day Earth-sized magnetosphere. By taking a surface planetary magnetic field similar to that of the Earth ($\sim 1$~G), we showed that the closest required orbital distance for a planet to have a magnetospheric size similar to that of the young-Earth decreases as compared to the present-day size of the magnetosphere  (Figure~\ref{fig.HZ}). While in the present-day scenario only two of our hypothetical planets would orbit inside the HZ, in the young-Earth scenario, $2/3$ of the planets would lie inside the HZ limits. 

\item Because stellar activity and rotation are related, we used in Section \ref{sec.results2} the empirically derived rotation-activity relation \citep{2009ApJ...692..538R} to investigate at which periods dM stars should be rotating such that if a terrestrial planet were found to orbit inside their HZ, the planet would still be able to sustain a magnetosphere similar to that of the Earth {(an analytical expression was provided in Eq.~(\ref{eq.analytic-prot}))}. We showed that early- and mid-dM stars (late-dM stars) with rotation periods longer than $37$ -- $202$~days ($\gtrsim 62$ -- $263$~days) would present large-scale magnetic fields that are small enough to not reduce the sizes of planet's magnetosphere to values below that of the Earth (Figure~\ref{fig.HZ_prot}). Because many late-dM stars probably rotate faster than this \citep{2008AJ....135..785W, 2011ApJ...727...56I}, conditions for terrestrial planets to harbour Earth-sized magnetospheres and larger are more easily achieved for planets orbiting slowly rotating early- and mid-dM stars.
\end{enumerate}
 
The particular study developed in this paper for the first time investigates the effects of a more magnetised environment surrounding a planet. We showed that to assess the extents of planetary magnetospheres (and, possibly, habitability) of planets orbiting dM stars, it is also important to understand the host-star magnetism, and in particular the large-scale stellar magnetic-field topology.  For example, although DX~Cnc and WX~UMa have similar masses, radii, rotation periods, and activity levels, they host very different large-scale magnetic fields (Table~\ref{table_B}). While a young-Earth-sized magnetosphere is possible for a planet in the HZ of DX~Cnc, the intense large-scale field strength of WX~UMa requires the planet to orbit outside the HZ for it to present a young-Earth-sized magnetosphere (Figure~\ref{fig.HZ}).
 
The small size of the sample adopted in Section \ref{sec.results} reflects the fact that large-scale magnetic fields have thus far been reconstructed only for a small number of stars in the M dwarf regime. An increased number of studied targets and a sample extended to even later spectral types ($\gtrsim$ M7), where magnetism is still poorly understood \citep{2010ApJ...710..924R} are desirable to advance studies like this one. To do this, new-generation instruments, such as SPIRou ({\bf sp}ectropolarim\`etre {\bf I}nfra{\bf rou}ge, PI: Donati), a near-infrared spectropolarimeter proposed for CFHT, will be fundamental. SPIRou will not only be capable of  simultaneously measuring both the large- and small-scale fields of dM stars, but will also be able to access a much larger sample of stars, including very inactive stars that are currently not accessible to ZDI. In addition,  SPIRou will also have stable radial velocity measurements, which are supposed to enable it to detect planets orbiting moderately active stars. As a result, when a planet is discovered with SPIRou, the information on the host star's magnetic field will be readily available.

\begin{acknowledgements} AAV acknowledges support from the Royal Astronomical Society through a post-doctoral fellowship. JM acknowledges support from a fellowship of the Alexander von Humboldt foundation. PL acknowledges funding from a STFC scholarship. AJBR is a Research Fellow of the Royal Commission for the Exhibition of 1851.
 \end{acknowledgements}

\let\astap=\aap
\let\apjlett=\apjl
\let\apjsupp=\apjs
\let\applopt=\ao
\let\mnrasl=\mnras

 \appendix
 \section{Model dependence on the source surface radius}\label{appendix}
 The position of the source surface is a free parameter in our model. We adopted $R_{\rm SS}=2.5R_\star$, motivated by solar observations and the study of the wind of the mid-M dwarf star V374~Peg \citep{2011MNRAS.412..351V}. Additional support for the choice of a small $R_{\rm SS}$ is found in the work of \citet{2000MNRAS.318.1217J}, who concluded that it may be difficult for active dM stars to sustain extended, quiescent coronae, especially for fast-rotating objects, with a small corotation radius. The reason is that at a certain height,  the closed loops no longer have the support of the local magnetic field, such that the hot X-ray material cannot any longer be restrained in the closed large-scale field and escapes in the form of a wind \citep{2005MNRAS.361.1173J,2006MNRAS.367..917J,2012MNRAS.421.1797A}. In addition, models of flaring loops in active dM stars suggest compact loop semi-lengths  \citep[$\lesssim 0.5R_\star$,][]{2000A&A...353..987F, 2000A&A...354.1021F, 2010ApJ...721..785O}
 
We also note that in the investigation of the coronal structure of the same sample of stars that was adopted in Section {\ref{sec.results}}, \citet{2012MNRAS.424.1077L} were able to reproduce the dependence of X-ray emission with Rossby number by adopting a source surface at $R_{\rm SS}=2.5R_\star$, giving us further confidence that the stars analysed here very likely have compact closed-field regions. 

Although physical arguments support our choice of small $R_{\rm SS}$, the source surface size is currently unconstrained, due to the lack of observations of stellar winds of dM stars. For this reason, we performed an analysis of the dependence of our results on the choice of $R_{\rm SS}$, noting that a larger source surface implies in a stronger decay of the stellar magnetic field for the closed-field corona. 

Adopting a larger radius for the source surface, we find that all our results remain qualitatively the same. Quantitatively, for $R_{\rm SS}=4~R_\star$, we find that 
(1) The largest magnetospheric sizes of planets with magnetisation like that of the Earth increased and  extended at most up to $7.4~r_p$, if they were orbiting at the inner edge of the HZ, or at most up to $14~r_p$ for outer edge-orbiters. Consequently, there is a decrease in the apertures of auroral ovals, now ranging from $21^\circ$ to $90^\circ$ for the inner edge of HZ and from $15^\circ$ to $45^\circ$ for the outer edge. 
(2) The closest orbital radius where Earth-sized magnetospheres would need to orbit is smaller on average by a factor $\sim1.3$ than the values reported in Figure~\ref{fig.HZ} and Table~\ref{table_B}. 
(3) The weakest  magnetic field required for a planet with a terrestrial magnetisation to sustain an Earth-sized magnetosphere if it orbits inside the HZ is smaller on average by a factor $\sim 1.7$ than the values of $B_{p,0}^{\min}$ reported in Table~\ref{table_B}. 
(4) The required stellar rotation rate for which a large-scale dipolar field would provide the same external pressure for a planet in the HZ of a dM star as the present-day value of the external pressure is $\gtrsim 29$ -- $155$~days for early- and mid-dM stars and $\gtrsim 48$ -- $202$~days for the late-dM stars. 
(5) The coefficient of Eq.~(\ref{eq.analytic-prot}) becomes $\mathcal{C}\simeq 13700$~days when we consider the same external pressure as the present-day value of the external pressure at the Earth, while for an external pressure similar to that of the young Earth,  $\mathcal{C}\simeq 3050$~days.

\listofobjects
\end{document}